\documentclass[reprint, superscriptaddress,prl]{revtex4-1}

\usepackage{graphicx}
\usepackage{dcolumn}
\usepackage{bm}
\usepackage{color} 
\usepackage[table]{xcolor}
\definecolor{Gray}{gray}{0.95}

\usepackage{soul}

\usepackage[normalem]{ulem}
\usepackage{color}

\begin{document}

\title{Charting Lattice Thermal Conductivity of Inorganic Crystals}

\author{Taishan Zhu}
\thanks{These two authors contributed equally.}
\author{Sheng Gong}%
\thanks{These two authors contributed equally.}
\affiliation{Department of Materials Science and Engineering, Massachusetts Institute of Technology, Cambridge, MA 20139
}%

\author{Tian Xie}%
\affiliation{Department of Materials Science and Engineering, Massachusetts Institute of Technology, Cambridge, MA 20139
}%

\author{Prashun Gorai}
\affiliation{Department of Metallurgical and Materials Engineering, Colorado School of Mines, Golden, CO 80401}

\author{Jeffrey C. Grossman}
\email[]{Corresponding author: jcg@mit.edu}
\affiliation{Department of Materials Science and Engineering, Massachusetts Institute of Technology, Cambridge, MA 20139
}%

\begin{abstract}
Thermal conductivity is a fundamental material property but challenging to predict, with less than 5\%  out of about $10^5$ synthesized inorganic materials being documented. In this work, we extract the structural chemistry that governs lattice thermal conductivity, by combining graph neural networks and random forest approaches. We show that both mean and variation of unit-cell configurational properties, such as atomic volume and bond length, are the most important features, followed by mass and elemental electronegativity. We chart the structural chemistry of lattice thermal conductivity into extended van-Arkel triangles, and predict the thermal conductivity of all known inorganic materials in the Inorganic Crystal Structure Database. For the latter, we develop a transfer learning framework extendable for other applications.
\end{abstract}

\maketitle

\section{Introduction}

Solids with both high and low extreme thermal conductivity have been pursued fundamentally and practically for decades  \cite{mukhopadhyay2018two,chiritescu2007ultralow,chen2020ultrahigh,li2018high,tian2018unusual,kang2018experimental,lindsay2013first,cahill2003nanoscale}, with diverse applications ranging from thermoelectrics \cite{he2017advances,nrm2017}, to thermal management in electronics {and avionics} \cite{schelling2005managing}, to high-temperature coatings in turbines \cite{clarke2005thermal} and human healthcare \cite{peng2020advanced}, to name only a few examples. Currently the records are held by diamond ($\sim$2000 W/mK) \cite{wei1993thermal} in the upper limit and aerogels ($\sim$0.01 W/mK) on the lower end \cite{lu1992thermal}, although it remains unclear whether these are hard limits. Regardless, the search for alternative materials that lie at or beyond these extremes is also of practical importance, particularly when multiple constraints are imposed, such as specific mechanical properties for thermal coatings \cite{clarke2005thermal} and (opto-) electronic properties for applications in energy conversion  \cite{he2017advances,zhu2019mixed}.

However, knowledge of the governing physics of {lattice} thermal conductivity ($\kappa$) remains incomplete at the atomic scale. \cite{toberer2011,seko2015prediction} 
Current understanding derives largely from kinetic theory and relates to unit cell properties (e.g., {symmetry}, {average} atomic {mass, volume, density}). \cite{lindsay2013first} 
{This understanding has been historically encapsulated into analytical models, in particular the Debye-Callaway (D-C) model \cite{morelli2006} and its extensions that incorporate the optical mode contributions.}\cite{toberer2011} Similarly, analytical models for $\kappa$ of solid-solution alloys, such as the Klemens model,\cite{gurunathan2020} are based on unit cell properties and scattering parameters. 
These models are explicit, but have parameters either numerically fitted or computed from first principles. 
For instance, Miller \textit{et al.} developed a modified D-C model with speed of sound and Gr\"{u}neisen parameter, which are derived from bulk modulus and average coordination number. \cite{miller2017capturing}

An emerging approach has been driven by learning from the existing data of $\kappa$, benefited from the developments in high-throughput screening and machine learning. \cite{Toher2014PRB,carrete2014finding, van2016high,Wang2011PRX,nrm2017} 
Through high-throughput calculations, databases are growing in size \emph{via} approaches for computing $\kappa$ based on Green-Kubo formalism  \cite{mcgaughey2006phonon,schelling2002comparison} and Boltzmann theory \cite{lindsay2013first, mcgaughey2019phonon}.
However, relying on dynamical and/or large-scale first-principles calculations, these methods are often computationally expensive, {and most high-throughput studies are thus far limited within certain material families.} \cite{van2016high} 
Alternatively, the above-mentioned semi-empirical models has also been successfully implemented for high-throughput predictions.\cite{gorai2016} 
Experimental data is even less available. To date, only some hundreds of the total $\sim 10^5$ synthesized materials documented in the Inorganic Crystal Structure Database (ICSD, 92919 {ordered} entries) have $\kappa$ values measured. \cite{gaultois2013} Thus, while machine learning techniques have shown initial success,  \cite{WEI2018908, seko2015prediction, ju2017designing, chen2019machine, kautz2019machine} both more data and novel approaches are needed in order to explore the vast materials space. 

\begin{figure*}
\includegraphics[width=0.95 \linewidth]{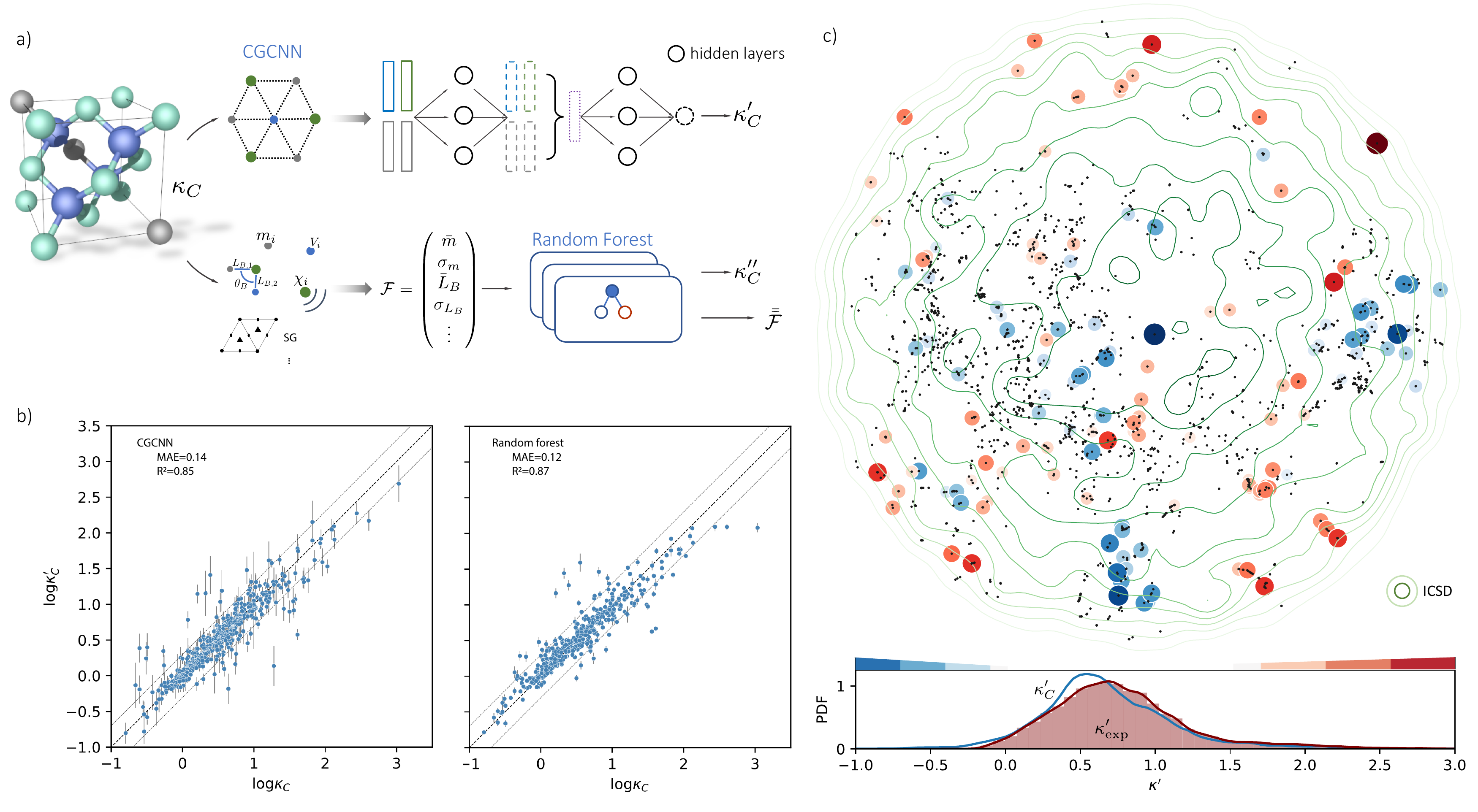}
\caption{\label{fig1}
(a) Schematic of two complementary models: CGCNN and random forest. (b) Predicted $\kappa_C'$ from these two models. The dashed band denotes a factor of 2. (c) High-throughput $\kappa_C'$ for all {ordered} ICSD structures, full data available online \cite{github_ML_kappa}. The contour denotes the distribution of ICSD materials in the feature space reduced to 2D \textit{via} PCA/t-SNE, along with the training set denoted by the dots. The histograms are the distribution of predicted $\kappa_C'$ and $\kappa_{exp}'$. See text for the prediction of $\kappa_{exp}'$. 
}
\end{figure*}

Towards this end, general guidelines for navigating and sampling the materials space for $\kappa$ will be valuable. Existing works for predicting/understanding $\kappa$ exhibit a catch-22 situation. 
On the one hand, descriptor-based methods assume \emph{a priori} knowledge of the physics of $\kappa$, so that appropriate features could be populated for materials. \cite{WEI2018908} However, since structural chemistry of $\kappa$ is largely unknown, the choice of atomic features is currently arbitrary. \cite{WEI2018908} On the other hand, techniques based on {graph neural networks} assume little pre-knowledge of $\kappa$, and can predict material properties directly from structure. \cite{xie2018crystal}
However, these methods must be utilized as ``black-boxes" \cite{rudin2019stop}, and the challenge of interpreting structure-$\kappa$ relation remains.

In this work, we predict $\kappa$ of all ordered and stoichiometric materials in ICSD, and then reveal the structural chemistry of $\kappa$. Two complementary approaches, neural network and random forest, are thus combined.
While the former is able to predict $\kappa$ directly from structures with little need for featurization, the latter allows us to extract the hidden chemistry in the dataset. With resolved important atomic and structural features that govern $\kappa$, we are able to chart the structural chemistry of $\kappa$ using our generalized van-Arkel triangles. 
{The training set is the calculated lattice thermal conductivity documented in TEDesignLab} ($\kappa_C$). \cite{gorai2016,miller2017capturing} Aiming at learning and predicting $\kappa$ measured by experiments, we also build an experimental dataset ($\kappa_{exp}$) collected from the literature (132 entries, available in SI).
To synergize the two datasets of varying size and with different fidelity, we extend our earlier graph neural networks model \cite{xie2018crystal} with transfer learning.

\begin{table}[b]
\caption{\label{tab_dft}
The predicted candidates in the lower and upper limits. $\kappa_{DFT}$, $\kappa_{C}$, are calculated values from DFT, and modified Debye-Callaway model \cite{miller2017capturing}, $\kappa_{C}'$ is the prediction from CGCNN, and $\kappa_{exp}'$ is the predicted experimental value with accompanying error  $\sigma_{\kappa_{exp}'}$. Note that $\kappa_{exp}'$ is from a  random  forest  model for the low  regime of $\kappa$, TL-CGCNN is used for high values. The unshaded materials are screened from ICSD, while the shaded materials are ``designed'' materials learnt from machine learning. The online database is an extended table with the random forest model. The standard deviation is given for $\log \kappa_{exp}'$.
}
\begin{ruledtabular}
\begin{tabular}{lcccccc}
 &$\log \kappa_{DFT}$ & $\log \kappa_{exp}$ &$ \log \kappa_{C}$ & $\log \kappa_{C}'$  &$\log \kappa_{exp}'$ & $\sigma$ \\
\hline
Cs$_2$BiAgCl$_6$ & -1.2 & - &  -0.06  & -0.1 & -0.3 & 0.1\\

CsTlF$_3$        & -1.0 & - &  0.01  & 0.2  & -0.1 & 0.1 \\ 

\rowcolor{Gray}
CsTlI$_3$        & -1.3 & - &  -0.25  & -0.1 & -0.3 & 0.0  \\ 

Tl$_3$VSe$_4$  & -0.8\cite{mukhopadhyay2018two} \footnotemark[1] & -0.5 \cite{mukhopadhyay2018two} & 0.03  & -0.2  & -0.3  &  0.0 \\

CsPbI$_3$ & -1.0 \cite{zhu2019mixed} & -0.4 \cite{lee2017ultralow} & -0.24  & -0.2 & -0.2 & 0.0\\

Be$_2$C & 2.06 &- & 1.94  & 2.9 & 2.6 & 0.4 \\

\rowcolor{Gray}
C$_3$N$_4$ & 2.4  & - & 2.66  &  2.5 & 2.6 & 0.2 \\

BP & 2.82 \cite{lindsay2013first} & 2.60 \cite{kumashiro1989thermal} & 2.45  & 2.4 & 2.6 & 0.2 \\

BAs & 3.50 \cite{lindsay2013first} & 3.08 \cite{li2018high,tian2018unusual,kang2018experimental} &  1.96 & 2.0 & 2.2 & 0.1 \\

BN & 3.33 \cite{lindsay2013first} & 3.20  \cite{chen2020ultrahigh} & 2.78 & 2.6 & 2.9 & 0.2 \\

Diamond & 3.54 \cite{lindsay2013first} & 3.36 \cite{wei1993thermal} & 3.01 & 3.1 & 3.4 & 0.3 \\ 
\end{tabular}
\end{ruledtabular}
\footnotetext[1]{New four-phonon plus SCPH calculations gives -0.5. }
\end{table}

\section{$\kappa$ for all known inorganic crystals}

We start by learning from our recently prepared high-throughput $\kappa_C$ dataset, \cite{gorai2016} before moving to the broader ICSD and the underlying structural chemistry. 
{The $\kappa_C$ dataset contains computed $\kappa$ of $\sim$2700 ordered and stoichiometric inorganic structures from the ICSD that have \textless50 atoms in the primitive cell and comprise mainly oxides (O), chalcogenides (S, Se, Te), and pnictides (N, P, As, Sb, Bi) - chemistries that are common among thermoelectric materials.} 
The predicted $\kappa$ are fairly accurate, with an average factor difference of 1.5 from experimentally measured values, over a range of $\kappa$ values that span 4 orders of magnitude.\cite{miller2017capturing} 
In this section, we will show both the transferability and limitation of this dataset, and in the next section we will show its implicit physics. 

Note that these two purposes suit two separate but complementary machine learning models: crystal graph convolutional neural network (CGCNN) \cite{xie2018crystal}, and interpretable random forest. 
These models are illustrated in Fig. \ref{fig1}(a), with further details available in the SI. Note that, instead of directly using the experimental dataset $\kappa_{exp}$, we choose our high-throughput dataset ($\kappa_C$) for these two purposes here because of the accuracy of $\kappa_C$, \cite{gorai2016} and also it is bigger in size than the $\kappa_{exp}$ dataset. 
For our high-throughput dataset, we randomly reserve 20\% entries as the test set, as plotted in Fig. \ref{fig1} (b). 
Both CGCNN and random forest models could predict $\log \kappa_C'$ with MAE$<0.15$ and $R^2>0.85$. 
When applied to our high-throughput dataset, these two methods are close in performance.

Moreover, different from CGCNN, random forest requires featurization for crystal structures before running through decision trees, which is largely physics-based and in many cases \emph{ad hoc}. 
Guided by lattice dynamical theory, we choose configurational features from elemental to atomic packing and bonding nature, which are constructed through Matminer \cite{ward2018matminer}, Magpie \cite{ward2016general}, and in-house codes. Since $\kappa$ is sensitive to both absolute values and variations of atomic properties, our feature engineering leads to a 154-dimensional descriptor, including the statistics (mean $\bar{.}$, standard deviation $\sigma .$, range $\{ . \}$ and mode) of {atomic number}, covalent radius ($r_a$), {atomic mass} ($m$), periodic table group and row number, Mendeleev number, volume per atom from ground state ($V_{GS}$), Pauling electronegativity ($\chi_a$), melting point ($T_m$), number ($N_V$) and unfilled ($N_U$) valence electrons in the s, p, d, and f shells of constituting elements, as well as structural features at the cell scale (space group, volume per atom $V_a$, packing fraction $\phi$, density $\rho$, bond length $L_B$,  bond angle $\theta_B$, and coordination number $CN$). The feature space spanned by these features is 154-dimensional, and we will show that this basis-choice is a good approximation in the next section, at least for the dataset currently available.

To visualize the feature space, we project it onto two dimensions, as shown in Fig. \ref{fig1}(c) (See also methods for dimension reduction in SI). Materials from our high-throughput dataset and the ICSD dataset are considered together, denoted by the contour. Most materials are populated in the central area, and the distribution varies smoothly. Our high-throughput entries, which are explicitly shown as scatter points, 
with the highest and lowest $\kappa$ values also highlighted, samples the reduced feature space quite satisfactorily in terms of uniformity. This suggests the {potential} transferability of our high-throughput dataset to ICSD. We did so using both CGCNN and random forest models, and we have made the data of $\kappa_C'$ available online \cite{github_ML_kappa}. As a first validation, we compare our predictions with experimental values. As shown in Fig. S1, 63(88) and 66(86) of 132 measured values align with our predictions within a factor of 2(3), for random forest and CGCNN, respectively. More detailed accuracy analysis, compared to different approaches, is presented in Tab. S3. We note that the accuracy is lower than the existing models (e.g. high-throughput \cite{miller2017capturing}), but our models predict $\kappa$ directly from atomic structure, without the need of expensive calculations for bulk modulus and Gr\"{u}neisen parameter. Instead, if we introduce bulk modulus into our random-forest model, we could reduce the MAE to $0.04$, which suggests the accuracy of our machine learning models could be at par with DFT predictions.
From the histogram in \ref{fig1}(c), the distribution of predicted $\kappa_C'$ follows approximately a normal distribution, with mean $\log \kappa \sim 0.8$ ($\bar{\kappa} \sim 6$W/mK) and standard deviation  $\sigma_{\log \kappa} \sim 0.5$.

To further validate our machine-learning predictions, we compare them to experimental measurements if available, and/or to first-principles calculations otherwise (Ref. \cite{phono3py}, see details in SI). The comparisons are presented in Table \ref{tab_dft} for several low- and high-$\kappa$ materials. The shaded entries are new materials suggested by machine learning, which are absent from ICSD and will be discussed later. Overall, our machine learning models can unanimously screen the lowest from the highest, which might be already sufficient for many materials selection/design scenarios, such as for thermoelectrics and thermal management, where either the lowest or the highest $\kappa$ values are sought. The other reason that we test our machine-learning models with these extremes is to show their reliability for extrapolation (transferability), which is often more challenging numerically than interpolation. Note that when we evaluate the testing materials, we removed them from the training set. For instance, diamond will be absent in the training set if diamond is being evaluated.

More quantitatively, the error of our machine learning models is comparable to first-principles calculations based on DFT. 
For instance, in the case of diamond, the extrapolated values, $\log \kappa=3.1$ and 3.4, are close to the experimental value 3.36, comparing to 3.54 obtained from DFT calculations. Such level of error is found to be applicable to all examined entries, except several outlying cases, such as BAs, for which the accuracy is less satisfactory, still 87\% data points are within a factor of 2. Other possible outliers are also observed when experimental values are missing and a substantial difference can be seen between DFT and machine learning, such as CsTlF$_3$ in Tab. \ref{tab_dft}. However, such possible outliers should be further examined (experimentally preferred) due to the possible underestimation from DFT calculations. In some cases, a difference of 50 - 100\% between DFT and experimental values can arise from the relaxation-time approximation up to 3-phonon interactions, which might be resolved by more sophisticated calculations, such as four-phonon and thermal-dependent dispersion. \cite{lindsay2019perspective, mukhopadhyay2018two, xia2020particlelike} In many other cases, our machine learning prediction can be even more accurate than DFT, such as the iodide perovskite CsPbI3 and the recently studied Tl$_3$VSe$_4$ (see Tab. \ref{tab_dft}). Moreover, note that our above error analyses is based on extrapolation. Even for the highest and lowest values, the machine learning models show satisfactory stability and prediction accuracy. For intermediate $\kappa$ values, machine learning models have better numerical performance due to their interpolative nature.

Note also that our predictions are directly mapped from atomic structures, without the need of any expensive atomistic calculations. In the case of diamond, the acceleration is 463,000s (DFT) versus $\sim 5$s (machine learning) using the same machine. Considering the high symmetry and small number of electrons/atoms in diamond, this acceleration rate could be readily exceeded, particularly for complex materials.

\begin{figure*}
\includegraphics[width=1.0 \linewidth]{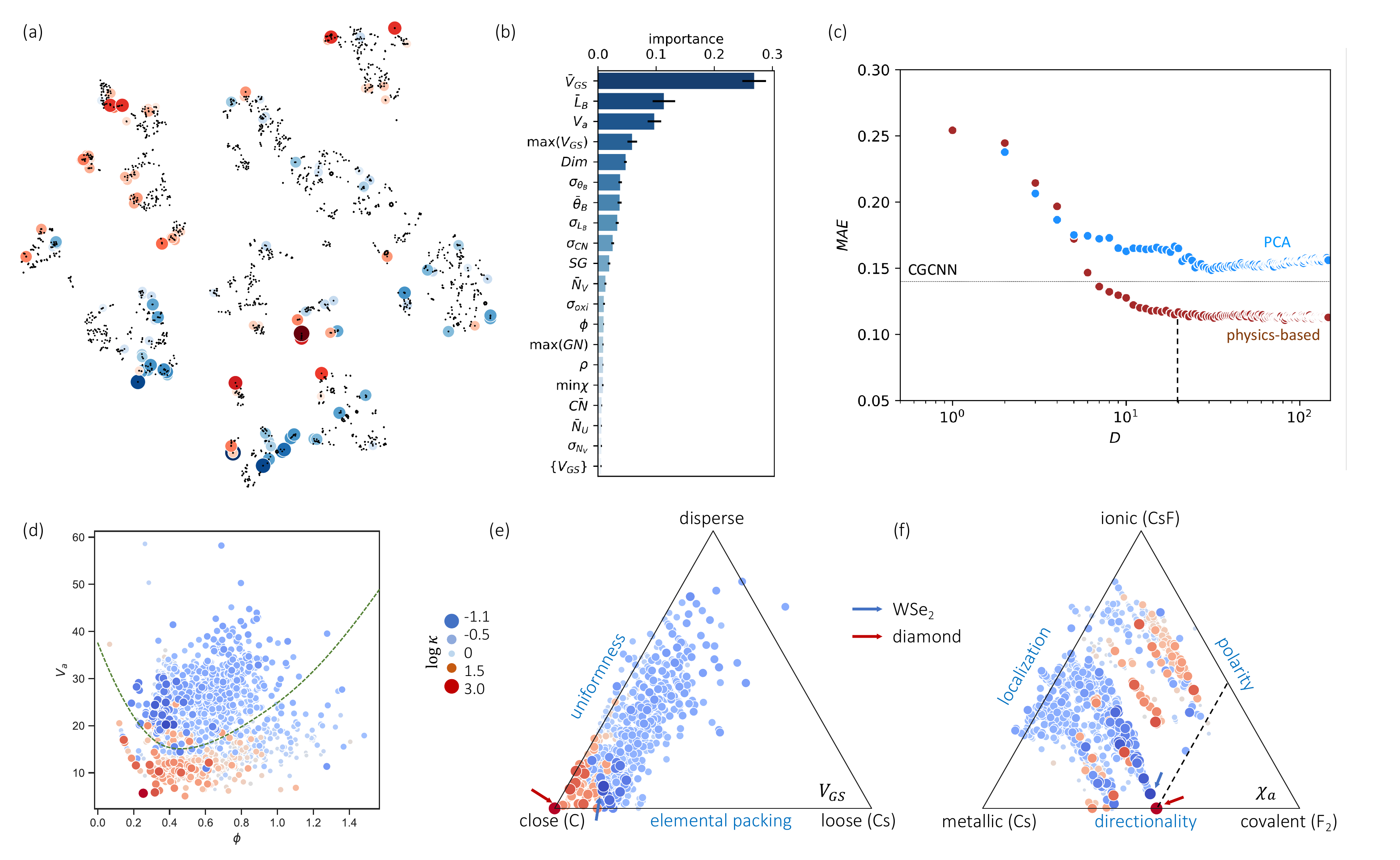}
\caption{\label{fig2}
(a) Clustering of the high-throughput database using PCA and tSNE, low-$\kappa$ and high-$\kappa$ entries are highlighted. (b) Top 20 important features and their F scores. (c) Dimension reduction by random-forest-ranked feature selection lead to even lower than PCA, and MAE approaches to CGCNN around 10 atomic features. Low-$\kappa$ and high-$\kappa$ materials can be divided by important features, (d) is an example of using $\phi-V_a$. (d-e) Chemical space illustrated by van-Arkel triangles, examples of structural ($V_{GS}$) and bonding ($\chi_a$) information. Similar triangles are available in Fig. S4. 
}
\end{figure*}

It is worth pointing out several observed limitations. Common to all machine learning approaches, these limitations result from finiteness of our dataset. For instance, although the projection of feature space to 2D in Fig. \ref{fig1}(c) shows uniform sampling, how it behaves in the 154-dimensional space should be further characterized. However, since the training set used is the largest reliable dataset available, this limitation will be translated to guidelines for future high-throughput calculations. This will be discussed further below when we extend to predicting experimental values $\kappa_{exp}'$. Moreover, the top 50 lowest-$\kappa$ and highest-$\kappa$ values are uniformly scattered, suggesting little knowledge content. However, as we present in Fig. \ref{fig2}(a), they are clustered when we plot without ICSD. This is another indication of the limited transferability to ICSD, but also demonstrates the knowledge content in our known dataset. 

\begin{figure*}
\includegraphics[width=0.75 \linewidth]{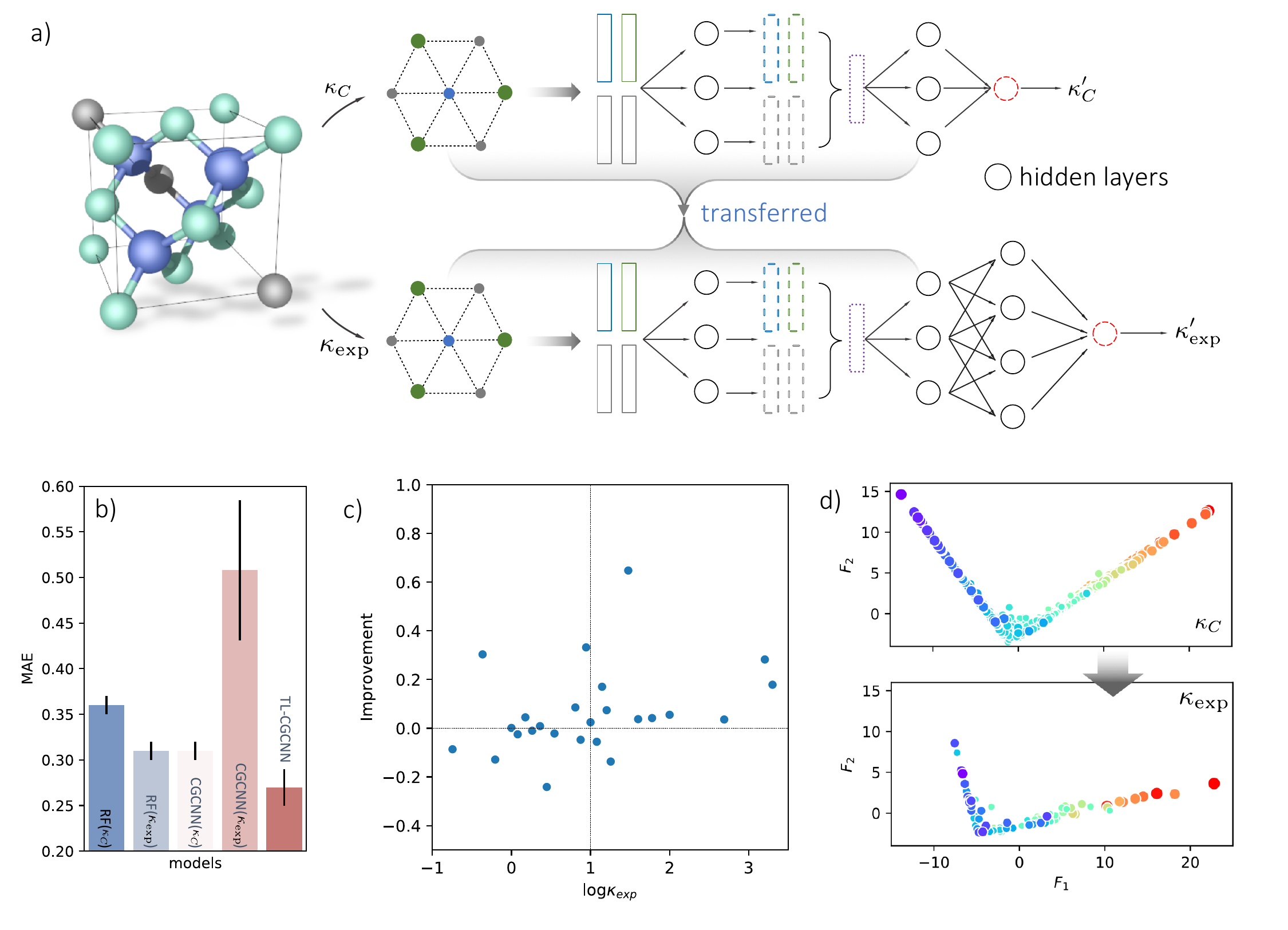}
\caption{\label{fig3}
(a) Transfer learning based on CGCNN (TL-CGCNN). This model learns high-throughput dataset $\kappa_C$ and transfer the knowledge to learning $\kappa_{exp}$. (b) Comparison between different machine learning models, including random forest, CGCNN, and TL-CGCNN, training on $\kappa_C$ or $\kappa_{exp}$.
TL-CGCNN exhibits the lowest MAE.
(c) A closer look at the improvement of TL-CGCNN compared with CGCNN($\kappa_C$) in prediction on the test set.
The region of $\log \kappa>1$ is systematically enhanced, while the $\log \kappa<1$ region can be better or worse.
To analyze the possible origin of this behavior, (d) plots the distribution of the feature space $\mathbf{V}_f$ projected onto two dimensions. 
The distribution and ranking of $\kappa_C$ is generally smoother than $\kappa_{exp}$, and for $\kappa_{exp}$ the upper end is smoother than the lower end.
}
\end{figure*}

\section{Structural chemistry of $\kappa$}

Such knowledge content in our dataset can be extracted in the form of ranked features (details in SI). In Fig. \ref{fig2}(b), the top 20 features are ranked in reducing order. These features include the elemental type ($V_{GS}$, $N_V$, $N_U$, $m$) and structural type. The latter consists of bonding properties ($L_B$, $\theta_B$, $CN$), and packing properties ($V_a$, $Dim$, $\phi$, $SG$, $\rho$). The learning of important features is different from a simple correlation relation (see Fig. S3). Fig. \ref{fig2}(c) shows the MAE as a function of increasing number of features, picking from the most important features, from PCA and random forest respectively. As the number of features increases, MAE reduces quickly and reaches our CGCNN accuracy with less than 10 features, and both are lower than PCA. The latter is usually chosen when little pre-knowledge is assumed, and our case shows that such purely data-driven techniques (e.g. PCA for dimensional reduction) could be excelled over by physics-informed approaches. Another interesting application of these important features is to physically categorize/cluster all the training materials. An example is shown in Fig. \ref{fig2}(d), where high-$\kappa$ and low-$\kappa$ values could be separated by the dashed line.

Further, we find that the mean-variance pair can be used to chart $\kappa$. Phonon transport is sensitive to chemical variations,
more than corresponding mean fields. Examples are mass and bond strength: the mean values define mean-field harmonic properties(e.g., group velocity), while the differences determine both harmonic (e.g., phononic bandgap) and anharmonic properties (e.g., higher-order force constants). This is also suggested in Fig. \ref{fig2}(b), where both mean values and variances are ranked top, such as $L_B$, $\theta_B$, $CN$, and  $N_V$.

Inspired by various forms of van-Arkel-type triangles, we use mean and standard deviation to construct extended triangles and generalize extensively to other atomic features. Invented originally for binary inorganic compounds, van-Arkel-type triangles were historically constructed to characterize the bonding nature, using the average and difference of the two elements' electronegativity $\chi_a$. In our case, we have multi-component compounds and more dominant quantities than $\chi_a$. Therefore, we extend the original van-Arkel triangle to include more components with mean and standard deviation, and to more physical descriptors important for $\kappa$. For instance, the $V_{GS}-$ and $\chi_a-$ triangles shown in Fig. \ref{fig2} (d-e) characterize packing and bonding information, respectively. More such charts are shown in Fig. S4. Although the extension is straightforward, it helps to chart the structural chemistry of $\kappa$. For instance, each of these triangles illustrates a projected materials space, all materials should be confined within these triangles, and the dots are our training set. While the coverage is essential for validating our dataset, it is also interesting to note that many of the chosen features are effective divisors (e.g. $V_{GS}$, $\chi_a$, $L_B$, $\theta_B$, $r_a$, $N_V$, $m$). In other words, given the mean and deviation of any of these features for a unit cell, the relative magnitude of $\kappa$ can already be estimated, at least qualitatively.

Note that our work based on physics 
and random forest confirms and also may enhance our existing understanding of trends in $\kappa$. For instance, it is commonly established that high-$\kappa$ materials often have i) low average atomic mass $\bar{m}$ (Fig. S4(g)), and ii) strong interatomic bonding, so that group velocity can be high, and iii) low anharmonicity in order to have large relaxation time (e.g. less scattering channels resulting from simple crystal structures). However, bonding strength and anharmonicity are computationally expensive quantities. Meanwhile, predicting $\kappa$ by studying solely the atomic structure was at best qualitative in the literature. With our analysis based on Figs. \ref{fig2} and S4, we now have proxies for bond strength and even $\kappa$, such as use $V_{GS}$,  $L_B$, and $r_a$. On the other hand, our analysis also shows that $\chi_a$ and $CN$ are more complicated than their reported influences. For instance, a strong correlation has been identified between $CN$ and $\kappa$. \cite{isaacs2020inverse} 

However, Fig. S4 (c) exhibits a rather mixed trend. This mixed trend for $CN$ can be understood by its competing impacts: i) higher $CN$ suggests larger anharmonicity due to more complexity in the bonding environment, ii) higher $CN$ means weaker bond strength, as stated by Pauling's second rule, due to electrostatic repulsion; iii) a large $CN$ also suggests stiff lattice, thus large sound speed. Therefore, the classic $\chi_a$ and $CN$ may only be sub-optimal features.

The structural chemistry of $\kappa$ can be used to extend the predictions from machine learning. For instance, in the upper limit, machine learning predicts the $\kappa$ values for BN and diamond to be 764 W/mK and 2225 W/mK, 
which are close to experimental values. As shown in Fig. S5(a), from the van-Arkel triangle of $\chi_a$, we notice two candidate materials between BN and diamond: C$_3$N$_4$ and B$_4$C$_3$. The $\kappa_{exp}'$ of C$_3$N$_4$ ranked in the top 1\% in our machine learning predictions over ICSD. In contrast, B$_4$C$_3$ is absent from ICSD, and is obtained by reading the van-Arkel triangle. One can also use this approach to search for low-$\kappa$ materials. Guided by the triangles, we adapt the corner of thalium, and iodine, considering their atomic weight and electronegativity. As shown in Fig. S5(b), binary and ternary compounds (e.g. TlI, CsTlF$_3$, CsPbI$_3$) are predicted from machine learning. Based on these, we could hypothesize that CsTlI$_3$ would have a low $\kappa$, which is also absent from the ICSD and confirmed by our DFT calculations (Tab. \ref{tab_dft}). 


\section{from $\kappa_C'$ to $\kappa_{exp}'$}

Thus far we have extended our $\kappa_C$ data to the ICSD set and learned information related to its structural chemistry. The next step is to use it to predict experimental measurements $\kappa_{exp}'$ directly. As mentioned, theoretical models are invaluable in that they contains the knowledge of $\kappa$, but due to simplicity they  inevitably can only provide insufficient accuracy and limited universality (e.g. the AFLOW dataset, see Fig. S1). Directly predicting experimental values has been attempted here using random forest and CGCNN, but with high MAEs (see Fig. \ref{fig3} (a) and Tab. S1), due to small size of the experimental dataset, $< 10^3$ entries. Note that this work predicts $\kappa$ also directly from atomic structure, without the need for bulk moduli and gruneisen parameters. This explains our higher MAE comparing to Refs \cite{miller2017capturing} and \cite{chen2019machine}. If we add bulk modulus into our dataset, we obtain an MAE of 0.04, far lower than the existing literature, which is lower than the existing literature. We aim to learn and predict $\kappa$ from the unit cell structure.

To take advantage of the knowledge learned from our larger high-throughput dataset, we develop a transfer learning framework demonstrated in Fig. \ref{fig3}(a).
This is a two-step CGCNN model: i) training a CGCNN model on our high-throughput $\kappa_C$ dataset, which has been done above.
ii) transferring the parameters of all layers from step i) to initialize a second CGCNN, and add one extra layer before the output layer.
For the second step, we use a smaller $\kappa$ dataset (132 entries) we collected from experimental measurements in the literature (see details in SI).
In step ii), all the layers other than the last one are frozen to keep the pre-learned knowledge and reduce the degrees of freedom to prevent overfitting. 
The overall performance is compared with random forest and CGCNN in Fig. \ref{fig3} (b), using different training datasets, and as can be seen our TL-CGCNN leads to the lowest MAE.
Figure \ref{fig3}(c) plots the improvement for each data in the test set, defined by the absolute error difference between CGCNN and TL-CGCNN.
It can seen that the accuracy on the high-$\kappa$ end ($\log \kappa>1$) is improved, but the accuracy is deteriorated on the low-$\kappa$ end, even though the overall performance is enhanced. 
Some example predictions in the high-$\kappa$ limit from step ii), termed $\kappa_{exp}'$, can be found in Tab. \ref{tab_dft}.
In the $\log \kappa<1$ region, we recommend $\kappa_{exp}'$ from random forest.

To understand the different performance in the high- and low- $\kappa$ regions of the transfer learning model, we look into the space of crystal features in the neural networks.
In Fig. \ref{fig3} (a), the network before the last hidden layer learns the feature vectors of materials $\mathbf{V}_f$, and the last operation from $\mathbf{V}_f$ to output is simply a regression with softmax activation. 
Since in TL-CGCNN we freeze $\mathbf{V}_f$ and all layers before the extra layer due to the limited amount of data, we essentially  use a one-layer neural network to fine tune $\kappa_{exp}$ learnt from $\kappa_{C}$. 
We plot $\mathbf{V}_f$ from the high-throughput and experimental datasets in Fig. \ref{fig3}(d). 
Interestingly, we observe a similar distribution between $\kappa_{exp}$ and $\kappa_C$ in the $\mathbf{V}_f$ space, showing a strong correlation between two datasets.
However, 
in the high-$\kappa$ region, $\kappa_{exp}$ distributes more smoothly along the V-shape than in the low-$\kappa$ region, which explains why TL-CGCNN performs better in high-$\kappa$ end. 
Such issue in the low-$\kappa$ region can be tackled from two aspects: 
i) more experimental data with low $\kappa$ should be generated to better understand $\kappa_{exp}$ distribution.
ii) future high-throughput calculations should be refined to shrink the difference between $\kappa_C$ and $\kappa_{exp}$, especially the outliers, in order to better sample the experimental $\mathbf{V}_f$ space.
the observation of data bias indicates the necessity of expanding the current database.
Instead of calculating hundreds of candidates in a certain material family each time, feature-space-based sampling techniques may be more computationally efficient to cover the material space.

In summary, we studied the structural chemistry of $\kappa$ for inorganic crystals, and predicted $\kappa$ for a large set of inorganic compounds, directly from their atomic structures.
We extended our graph neural network model to inclcude transfer learning, and using as input our recently prepared database of lattice conductivity $\kappa$.
Combining the neural networks model and interpretable random forest, we extract atomic features that dominate the physics of $\kappa$, including elemental ($\chi_a$, $V_{GS}$, $r_a$) and packing ($L_B$, $V_a$).
Other features, such as $CN$, are shown to be also important but more complicated than conventionally assumed.
Using these important features, we propose feature-space sampling for future high-throughput coverage of materials space.
The combination of machine learning search and the learned structural chemistry sheds light on bottom-up design of materials from elements.

\begin{acknowledgments}
This work is supported by various computational resources: (i) Comet at the Extreme Science and Engineering Discovery Environment (XSEDE), which is supported by National Science Foundation grant number ACI-1548562, through allocation TG-DMR090027, and (ii) the National Energy Research Scientific Computing Center (NERSC), which is supported by the Office of Science of the U.S. Department of Energy under Contract No. DE-AC02-05CH11231. P. G. acknowledges support from the Advanced Research Projects Agency-Energy (ARPA-E), U.S. Department of Energy, under Award Number DE-AR0001205. The views and opinions of authors expressed herein do not necessarily state or reflect those of the United States Government or any agency thereof. The research was performed using computational resources sponsored by the Department of Energy's Office of Energy Efficiency and Renewable Energy and located at the NREL.
\end{acknowledgments}

\bibliography{kML}

\begin{thebibliography}{45}%
\makeatletter
\providecommand \@ifxundefined [1]{%
 \@ifx{#1\undefined}
}%
\providecommand \@ifnum [1]{%
 \ifnum #1\expandafter \@firstoftwo
 \else \expandafter \@secondoftwo
 \fi
}%
\providecommand \@ifx [1]{%
 \ifx #1\expandafter \@firstoftwo
 \else \expandafter \@secondoftwo
 \fi
}%
\providecommand \natexlab [1]{#1}%
\providecommand \enquote  [1]{``#1''}%
\providecommand \bibnamefont  [1]{#1}%
\providecommand \bibfnamefont [1]{#1}%
\providecommand \citenamefont [1]{#1}%
\providecommand \href@noop [0]{\@secondoftwo}%
\providecommand \href [0]{\begingroup \@sanitize@url \@href}%
\providecommand \@href[1]{\@@startlink{#1}\@@href}%
\providecommand \@@href[1]{\endgroup#1\@@endlink}%
\providecommand \@sanitize@url [0]{\catcode `\\12\catcode `\$12\catcode
  `\&12\catcode `\#12\catcode `\^12\catcode `\_12\catcode `\%12\relax}%
\providecommand \@@startlink[1]{}%
\providecommand \@@endlink[0]{}%
\providecommand \url  [0]{\begingroup\@sanitize@url \@url }%
\providecommand \@url [1]{\endgroup\@href {#1}{\urlprefix }}%
\providecommand \urlprefix  [0]{URL }%
\providecommand \Eprint [0]{\href }%
\providecommand \doibase [0]{https://doi.org/}%
\providecommand \selectlanguage [0]{\@gobble}%
\providecommand \bibinfo  [0]{\@secondoftwo}%
\providecommand \bibfield  [0]{\@secondoftwo}%
\providecommand \translation [1]{[#1]}%
\providecommand \BibitemOpen [0]{}%
\providecommand \bibitemStop [0]{}%
\providecommand \bibitemNoStop [0]{.\EOS\space}%
\providecommand \EOS [0]{\spacefactor3000\relax}%
\providecommand \BibitemShut  [1]{\csname bibitem#1\endcsname}%
\let\auto@bib@innerbib\@empty
\bibitem [{\citenamefont {Mukhopadhyay}\ \emph {et~al.}(2018)\citenamefont
  {Mukhopadhyay}, \citenamefont {Parker}, \citenamefont {Sales}, \citenamefont
  {Puretzky}, \citenamefont {McGuire},\ and\ \citenamefont
  {Lindsay}}]{mukhopadhyay2018two}%
  \BibitemOpen
  \bibfield  {author} {\bibinfo {author} {\bibfnamefont {S.}~\bibnamefont
  {Mukhopadhyay}}, \bibinfo {author} {\bibfnamefont {D.~S.}\ \bibnamefont
  {Parker}}, \bibinfo {author} {\bibfnamefont {B.~C.}\ \bibnamefont {Sales}},
  \bibinfo {author} {\bibfnamefont {A.~A.}\ \bibnamefont {Puretzky}}, \bibinfo
  {author} {\bibfnamefont {M.~A.}\ \bibnamefont {McGuire}},\ and\ \bibinfo
  {author} {\bibfnamefont {L.}~\bibnamefont {Lindsay}},\ }\bibfield  {title}
  {\bibinfo {title} {Two-channel model for ultralow thermal conductivity of
  crystalline {Tl$_3$VSe$_4$}},\ }\href@noop {} {\bibfield  {journal} {\bibinfo
   {journal} {Science}\ }\textbf {\bibinfo {volume} {360}},\ \bibinfo {pages}
  {1455} (\bibinfo {year} {2018})}\BibitemShut {NoStop}%
\bibitem [{\citenamefont {Chiritescu}\ \emph {et~al.}(2007)\citenamefont
  {Chiritescu}, \citenamefont {Cahill}, \citenamefont {Nguyen}, \citenamefont
  {Johnson}, \citenamefont {Bodapati}, \citenamefont {Keblinski},\ and\
  \citenamefont {Zschack}}]{chiritescu2007ultralow}%
  \BibitemOpen
  \bibfield  {author} {\bibinfo {author} {\bibfnamefont {C.}~\bibnamefont
  {Chiritescu}}, \bibinfo {author} {\bibfnamefont {D.~G.}\ \bibnamefont
  {Cahill}}, \bibinfo {author} {\bibfnamefont {N.}~\bibnamefont {Nguyen}},
  \bibinfo {author} {\bibfnamefont {D.}~\bibnamefont {Johnson}}, \bibinfo
  {author} {\bibfnamefont {A.}~\bibnamefont {Bodapati}}, \bibinfo {author}
  {\bibfnamefont {P.}~\bibnamefont {Keblinski}},\ and\ \bibinfo {author}
  {\bibfnamefont {P.}~\bibnamefont {Zschack}},\ }\bibfield  {title} {\bibinfo
  {title} {Ultralow thermal conductivity in disordered, layered {WSe$_2$}
  crystals},\ }\href@noop {} {\bibfield  {journal} {\bibinfo  {journal}
  {Science}\ }\textbf {\bibinfo {volume} {315}},\ \bibinfo {pages} {351}
  (\bibinfo {year} {2007})}\BibitemShut {NoStop}%
\bibitem [{\citenamefont {Chen}\ \emph {et~al.}(2020)\citenamefont {Chen},
  \citenamefont {Song}, \citenamefont {Ravichandran}, \citenamefont {Zheng},
  \citenamefont {Chen}, \citenamefont {Lee}, \citenamefont {Sun}, \citenamefont
  {Li}, \citenamefont {Gamage}, \citenamefont {Tian} \emph
  {et~al.}}]{chen2020ultrahigh}%
  \BibitemOpen
  \bibfield  {author} {\bibinfo {author} {\bibfnamefont {K.}~\bibnamefont
  {Chen}}, \bibinfo {author} {\bibfnamefont {B.}~\bibnamefont {Song}}, \bibinfo
  {author} {\bibfnamefont {N.~K.}\ \bibnamefont {Ravichandran}}, \bibinfo
  {author} {\bibfnamefont {Q.}~\bibnamefont {Zheng}}, \bibinfo {author}
  {\bibfnamefont {X.}~\bibnamefont {Chen}}, \bibinfo {author} {\bibfnamefont
  {H.}~\bibnamefont {Lee}}, \bibinfo {author} {\bibfnamefont {H.}~\bibnamefont
  {Sun}}, \bibinfo {author} {\bibfnamefont {S.}~\bibnamefont {Li}}, \bibinfo
  {author} {\bibfnamefont {G.~A.}\ \bibnamefont {Gamage}}, \bibinfo {author}
  {\bibfnamefont {F.}~\bibnamefont {Tian}}, \emph {et~al.},\ }\bibfield
  {title} {\bibinfo {title} {Ultrahigh thermal conductivity in isotope-enriched
  cubic boron nitride},\ }\href@noop {} {\bibfield  {journal} {\bibinfo
  {journal} {Science}\ } (\bibinfo {year} {2020})}\BibitemShut {NoStop}%
\bibitem [{\citenamefont {Li}\ \emph {et~al.}(2018)\citenamefont {Li},
  \citenamefont {Zheng}, \citenamefont {Lv}, \citenamefont {Liu}, \citenamefont
  {Wang}, \citenamefont {Huang}, \citenamefont {Cahill},\ and\ \citenamefont
  {Lv}}]{li2018high}%
  \BibitemOpen
  \bibfield  {author} {\bibinfo {author} {\bibfnamefont {S.}~\bibnamefont
  {Li}}, \bibinfo {author} {\bibfnamefont {Q.}~\bibnamefont {Zheng}}, \bibinfo
  {author} {\bibfnamefont {Y.}~\bibnamefont {Lv}}, \bibinfo {author}
  {\bibfnamefont {X.}~\bibnamefont {Liu}}, \bibinfo {author} {\bibfnamefont
  {X.}~\bibnamefont {Wang}}, \bibinfo {author} {\bibfnamefont {P.~Y.}\
  \bibnamefont {Huang}}, \bibinfo {author} {\bibfnamefont {D.~G.}\ \bibnamefont
  {Cahill}},\ and\ \bibinfo {author} {\bibfnamefont {B.}~\bibnamefont {Lv}},\
  }\bibfield  {title} {\bibinfo {title} {High thermal conductivity in cubic
  boron arsenide crystals},\ }\href@noop {} {\bibfield  {journal} {\bibinfo
  {journal} {Science}\ }\textbf {\bibinfo {volume} {361}},\ \bibinfo {pages}
  {579} (\bibinfo {year} {2018})}\BibitemShut {NoStop}%
\bibitem [{\citenamefont {Tian}\ \emph {et~al.}(2018)\citenamefont {Tian},
  \citenamefont {Song}, \citenamefont {Chen}, \citenamefont {Ravichandran},
  \citenamefont {Lv}, \citenamefont {Chen}, \citenamefont {Sullivan},
  \citenamefont {Kim}, \citenamefont {Zhou}, \citenamefont {Liu} \emph
  {et~al.}}]{tian2018unusual}%
  \BibitemOpen
  \bibfield  {author} {\bibinfo {author} {\bibfnamefont {F.}~\bibnamefont
  {Tian}}, \bibinfo {author} {\bibfnamefont {B.}~\bibnamefont {Song}}, \bibinfo
  {author} {\bibfnamefont {X.}~\bibnamefont {Chen}}, \bibinfo {author}
  {\bibfnamefont {N.~K.}\ \bibnamefont {Ravichandran}}, \bibinfo {author}
  {\bibfnamefont {Y.}~\bibnamefont {Lv}}, \bibinfo {author} {\bibfnamefont
  {K.}~\bibnamefont {Chen}}, \bibinfo {author} {\bibfnamefont {S.}~\bibnamefont
  {Sullivan}}, \bibinfo {author} {\bibfnamefont {J.}~\bibnamefont {Kim}},
  \bibinfo {author} {\bibfnamefont {Y.}~\bibnamefont {Zhou}}, \bibinfo {author}
  {\bibfnamefont {T.-H.}\ \bibnamefont {Liu}}, \emph {et~al.},\ }\bibfield
  {title} {\bibinfo {title} {Unusual high thermal conductivity in boron
  arsenide bulk crystals},\ }\href@noop {} {\bibfield  {journal} {\bibinfo
  {journal} {Science}\ }\textbf {\bibinfo {volume} {361}},\ \bibinfo {pages}
  {582} (\bibinfo {year} {2018})}\BibitemShut {NoStop}%
\bibitem [{\citenamefont {Kang}\ \emph {et~al.}(2018)\citenamefont {Kang},
  \citenamefont {Li}, \citenamefont {Wu}, \citenamefont {Nguyen},\ and\
  \citenamefont {Hu}}]{kang2018experimental}%
  \BibitemOpen
  \bibfield  {author} {\bibinfo {author} {\bibfnamefont {J.~S.}\ \bibnamefont
  {Kang}}, \bibinfo {author} {\bibfnamefont {M.}~\bibnamefont {Li}}, \bibinfo
  {author} {\bibfnamefont {H.}~\bibnamefont {Wu}}, \bibinfo {author}
  {\bibfnamefont {H.}~\bibnamefont {Nguyen}},\ and\ \bibinfo {author}
  {\bibfnamefont {Y.}~\bibnamefont {Hu}},\ }\bibfield  {title} {\bibinfo
  {title} {Experimental observation of high thermal conductivity in boron
  arsenide},\ }\href@noop {} {\bibfield  {journal} {\bibinfo  {journal}
  {Science}\ }\textbf {\bibinfo {volume} {361}},\ \bibinfo {pages} {575}
  (\bibinfo {year} {2018})}\BibitemShut {NoStop}%
\bibitem [{\citenamefont {Lindsay}\ \emph {et~al.}(2013)\citenamefont
  {Lindsay}, \citenamefont {Broido},\ and\ \citenamefont
  {Reinecke}}]{lindsay2013first}%
  \BibitemOpen
  \bibfield  {author} {\bibinfo {author} {\bibfnamefont {L.}~\bibnamefont
  {Lindsay}}, \bibinfo {author} {\bibfnamefont {D.}~\bibnamefont {Broido}},\
  and\ \bibinfo {author} {\bibfnamefont {T.}~\bibnamefont {Reinecke}},\
  }\bibfield  {title} {\bibinfo {title} {First-principles determination of
  ultrahigh thermal conductivity of boron arsenide: A competitor for
  diamond?},\ }\href@noop {} {\bibfield  {journal} {\bibinfo  {journal}
  {Physical Review Letters}\ }\textbf {\bibinfo {volume} {111}},\ \bibinfo
  {pages} {025901} (\bibinfo {year} {2013})}\BibitemShut {NoStop}%
\bibitem [{\citenamefont {Cahill}\ \emph {et~al.}(2003)\citenamefont {Cahill},
  \citenamefont {Ford}, \citenamefont {Goodson}, \citenamefont {Mahan},
  \citenamefont {Majumdar}, \citenamefont {Maris}, \citenamefont {Merlin},\
  and\ \citenamefont {Phillpot}}]{cahill2003nanoscale}%
  \BibitemOpen
  \bibfield  {author} {\bibinfo {author} {\bibfnamefont {D.~G.}\ \bibnamefont
  {Cahill}}, \bibinfo {author} {\bibfnamefont {W.~K.}\ \bibnamefont {Ford}},
  \bibinfo {author} {\bibfnamefont {K.~E.}\ \bibnamefont {Goodson}}, \bibinfo
  {author} {\bibfnamefont {G.~D.}\ \bibnamefont {Mahan}}, \bibinfo {author}
  {\bibfnamefont {A.}~\bibnamefont {Majumdar}}, \bibinfo {author}
  {\bibfnamefont {H.~J.}\ \bibnamefont {Maris}}, \bibinfo {author}
  {\bibfnamefont {R.}~\bibnamefont {Merlin}},\ and\ \bibinfo {author}
  {\bibfnamefont {S.~R.}\ \bibnamefont {Phillpot}},\ }\bibfield  {title}
  {\bibinfo {title} {Nanoscale thermal transport},\ }\href@noop {} {\bibfield
  {journal} {\bibinfo  {journal} {Journal of Applied Physics}\ }\textbf
  {\bibinfo {volume} {93}},\ \bibinfo {pages} {793} (\bibinfo {year}
  {2003})}\BibitemShut {NoStop}%
\bibitem [{\citenamefont {He}\ and\ \citenamefont
  {Tritt}(2017)}]{he2017advances}%
  \BibitemOpen
  \bibfield  {author} {\bibinfo {author} {\bibfnamefont {J.}~\bibnamefont
  {He}}\ and\ \bibinfo {author} {\bibfnamefont {T.~M.}\ \bibnamefont {Tritt}},\
  }\bibfield  {title} {\bibinfo {title} {Advances in thermoelectric materials
  research: Looking back and moving forward},\ }\href@noop {} {\bibfield
  {journal} {\bibinfo  {journal} {Science}\ }\textbf {\bibinfo {volume}
  {357}},\ \bibinfo {pages} {eaak9997} (\bibinfo {year} {2017})}\BibitemShut
  {NoStop}%
\bibitem [{\citenamefont {Gorai}\ \emph {et~al.}(2017)\citenamefont {Gorai},
  \citenamefont {Stevanovi{\'c}},\ and\ \citenamefont {Toberer}}]{nrm2017}%
  \BibitemOpen
  \bibfield  {author} {\bibinfo {author} {\bibfnamefont {P.}~\bibnamefont
  {Gorai}}, \bibinfo {author} {\bibfnamefont {V.}~\bibnamefont
  {Stevanovi{\'c}}},\ and\ \bibinfo {author} {\bibfnamefont {E.~S.}\
  \bibnamefont {Toberer}},\ }\bibfield  {title} {\bibinfo {title}
  {Computationally guided discovery of thermoelectric materials},\ }\href
  {https://doi.org/10.1038/natrevmats.2017.53} {\bibfield  {journal} {\bibinfo
  {journal} {Nature Reviews Materials}\ }\textbf {\bibinfo {volume} {2}},\
  \bibinfo {pages} {1} (\bibinfo {year} {2017})}\BibitemShut {NoStop}%
\bibitem [{\citenamefont {Schelling}\ \emph {et~al.}(2005)\citenamefont
  {Schelling}, \citenamefont {Shi},\ and\ \citenamefont
  {Goodson}}]{schelling2005managing}%
  \BibitemOpen
  \bibfield  {author} {\bibinfo {author} {\bibfnamefont {P.~K.}\ \bibnamefont
  {Schelling}}, \bibinfo {author} {\bibfnamefont {L.}~\bibnamefont {Shi}},\
  and\ \bibinfo {author} {\bibfnamefont {K.~E.}\ \bibnamefont {Goodson}},\
  }\bibfield  {title} {\bibinfo {title} {Managing heat for electronics},\
  }\href@noop {} {\bibfield  {journal} {\bibinfo  {journal} {Materials Today}\
  }\textbf {\bibinfo {volume} {8}},\ \bibinfo {pages} {30} (\bibinfo {year}
  {2005})}\BibitemShut {NoStop}%
\bibitem [{\citenamefont {Clarke}\ and\ \citenamefont
  {Phillpot}(2005)}]{clarke2005thermal}%
  \BibitemOpen
  \bibfield  {author} {\bibinfo {author} {\bibfnamefont {D.~R.}\ \bibnamefont
  {Clarke}}\ and\ \bibinfo {author} {\bibfnamefont {S.~R.}\ \bibnamefont
  {Phillpot}},\ }\bibfield  {title} {\bibinfo {title} {Thermal barrier coating
  materials},\ }\href@noop {} {\bibfield  {journal} {\bibinfo  {journal}
  {Materials Today}\ }\textbf {\bibinfo {volume} {8}},\ \bibinfo {pages} {22}
  (\bibinfo {year} {2005})}\BibitemShut {NoStop}%
\bibitem [{\citenamefont {Peng}\ and\ \citenamefont
  {Cui}(2020)}]{peng2020advanced}%
  \BibitemOpen
  \bibfield  {author} {\bibinfo {author} {\bibfnamefont {Y.}~\bibnamefont
  {Peng}}\ and\ \bibinfo {author} {\bibfnamefont {Y.}~\bibnamefont {Cui}},\
  }\bibfield  {title} {\bibinfo {title} {Advanced textiles for personal thermal
  management and energy},\ }\href@noop {} {\bibfield  {journal} {\bibinfo
  {journal} {Joule}\ } (\bibinfo {year} {2020})}\BibitemShut {NoStop}%
\bibitem [{\citenamefont {Wei}\ \emph {et~al.}(1993)\citenamefont {Wei},
  \citenamefont {Kuo}, \citenamefont {Thomas}, \citenamefont {Anthony},\ and\
  \citenamefont {Banholzer}}]{wei1993thermal}%
  \BibitemOpen
  \bibfield  {author} {\bibinfo {author} {\bibfnamefont {L.}~\bibnamefont
  {Wei}}, \bibinfo {author} {\bibfnamefont {P.}~\bibnamefont {Kuo}}, \bibinfo
  {author} {\bibfnamefont {R.}~\bibnamefont {Thomas}}, \bibinfo {author}
  {\bibfnamefont {T.}~\bibnamefont {Anthony}},\ and\ \bibinfo {author}
  {\bibfnamefont {W.}~\bibnamefont {Banholzer}},\ }\bibfield  {title} {\bibinfo
  {title} {Thermal conductivity of isotopically modified single crystal
  diamond},\ }\href@noop {} {\bibfield  {journal} {\bibinfo  {journal}
  {Physical Review Letters}\ }\textbf {\bibinfo {volume} {70}},\ \bibinfo
  {pages} {3764} (\bibinfo {year} {1993})}\BibitemShut {NoStop}%
\bibitem [{\citenamefont {Lu}\ \emph {et~al.}(1992)\citenamefont {Lu},
  \citenamefont {Arduini-Schuster}, \citenamefont {Kuhn}, \citenamefont
  {Nilsson}, \citenamefont {Fricke},\ and\ \citenamefont
  {Pekala}}]{lu1992thermal}%
  \BibitemOpen
  \bibfield  {author} {\bibinfo {author} {\bibfnamefont {X.}~\bibnamefont
  {Lu}}, \bibinfo {author} {\bibfnamefont {M.}~\bibnamefont
  {Arduini-Schuster}}, \bibinfo {author} {\bibfnamefont {J.}~\bibnamefont
  {Kuhn}}, \bibinfo {author} {\bibfnamefont {O.}~\bibnamefont {Nilsson}},
  \bibinfo {author} {\bibfnamefont {J.}~\bibnamefont {Fricke}},\ and\ \bibinfo
  {author} {\bibfnamefont {R.}~\bibnamefont {Pekala}},\ }\bibfield  {title}
  {\bibinfo {title} {Thermal conductivity of monolithic organic aerogels},\
  }\href@noop {} {\bibfield  {journal} {\bibinfo  {journal} {Science}\ }\textbf
  {\bibinfo {volume} {255}},\ \bibinfo {pages} {971} (\bibinfo {year}
  {1992})}\BibitemShut {NoStop}%
\bibitem [{\citenamefont {Zhu}\ and\ \citenamefont
  {Ertekin}(2019)}]{zhu2019mixed}%
  \BibitemOpen
  \bibfield  {author} {\bibinfo {author} {\bibfnamefont {T.}~\bibnamefont
  {Zhu}}\ and\ \bibinfo {author} {\bibfnamefont {E.}~\bibnamefont {Ertekin}},\
  }\bibfield  {title} {\bibinfo {title} {Mixed phononic and non-phononic
  transport in hybrid lead halide perovskites: glass-crystal duality, dynamical
  disorder, and anharmonicity},\ }\href@noop {} {\bibfield  {journal} {\bibinfo
   {journal} {Energy \& Environmental Science}\ }\textbf {\bibinfo {volume}
  {12}},\ \bibinfo {pages} {216} (\bibinfo {year} {2019})}\BibitemShut
  {NoStop}%
\bibitem [{\citenamefont {Toberer}\ \emph {et~al.}(2011)\citenamefont
  {Toberer}, \citenamefont {Zevalkink},\ and\ \citenamefont
  {Snyder}}]{toberer2011}%
  \BibitemOpen
  \bibfield  {author} {\bibinfo {author} {\bibfnamefont {E.~S.}\ \bibnamefont
  {Toberer}}, \bibinfo {author} {\bibfnamefont {A.}~\bibnamefont {Zevalkink}},\
  and\ \bibinfo {author} {\bibfnamefont {G.~J.}\ \bibnamefont {Snyder}},\
  }\bibfield  {title} {\bibinfo {title} {Phonon engineering through crystal
  chemistry},\ }\href {https://doi.org/10.1039/C1JM11754H} {\bibfield
  {journal} {\bibinfo  {journal} {Journal Materials Chemistry}\ }\textbf
  {\bibinfo {volume} {21}},\ \bibinfo {pages} {15843} (\bibinfo {year}
  {2011})}\BibitemShut {NoStop}%
\bibitem [{\citenamefont {Seko}\ \emph {et~al.}(2015)\citenamefont {Seko},
  \citenamefont {Togo}, \citenamefont {Hayashi}, \citenamefont {Tsuda},
  \citenamefont {Chaput},\ and\ \citenamefont {Tanaka}}]{seko2015prediction}%
  \BibitemOpen
  \bibfield  {author} {\bibinfo {author} {\bibfnamefont {A.}~\bibnamefont
  {Seko}}, \bibinfo {author} {\bibfnamefont {A.}~\bibnamefont {Togo}}, \bibinfo
  {author} {\bibfnamefont {H.}~\bibnamefont {Hayashi}}, \bibinfo {author}
  {\bibfnamefont {K.}~\bibnamefont {Tsuda}}, \bibinfo {author} {\bibfnamefont
  {L.}~\bibnamefont {Chaput}},\ and\ \bibinfo {author} {\bibfnamefont
  {I.}~\bibnamefont {Tanaka}},\ }\bibfield  {title} {\bibinfo {title}
  {Prediction of low-thermal-conductivity compounds with first-principles
  anharmonic lattice-dynamics calculations and bayesian optimization},\
  }\href@noop {} {\bibfield  {journal} {\bibinfo  {journal} {Physical Review
  Letters}\ }\textbf {\bibinfo {volume} {115}},\ \bibinfo {pages} {205901}
  (\bibinfo {year} {2015})}\BibitemShut {NoStop}%
\bibitem [{\citenamefont {Morelli}\ and\ \citenamefont
  {Slack}(2006)}]{morelli2006}%
  \BibitemOpen
  \bibfield  {author} {\bibinfo {author} {\bibfnamefont {D.~T.}\ \bibnamefont
  {Morelli}}\ and\ \bibinfo {author} {\bibfnamefont {G.~A.}\ \bibnamefont
  {Slack}},\ }\bibfield  {title} {\bibinfo {title} {High lattice thermal
  conductivity solids},\ }in\ \href@noop {} {\emph {\bibinfo {booktitle} {High
  Thermal Conductivity Materials}}}\ (\bibinfo  {publisher} {Springer},\
  \bibinfo {year} {2006})\ p.~\bibinfo {pages} {37}\BibitemShut {NoStop}%
\bibitem [{\citenamefont {Gurunathan}\ \emph {et~al.}(2020)\citenamefont
  {Gurunathan}, \citenamefont {Hanus},\ and\ \citenamefont
  {Snyder}}]{gurunathan2020}%
  \BibitemOpen
  \bibfield  {author} {\bibinfo {author} {\bibfnamefont {R.}~\bibnamefont
  {Gurunathan}}, \bibinfo {author} {\bibfnamefont {R.}~\bibnamefont {Hanus}},\
  and\ \bibinfo {author} {\bibfnamefont {G.~J.}\ \bibnamefont {Snyder}},\
  }\bibfield  {title} {\bibinfo {title} {Alloy scattering of phonons},\ }\href
  {https://doi.org/10.1039/C9MH01990A} {\bibfield  {journal} {\bibinfo
  {journal} {Materials Horizon}\ }\textbf {\bibinfo {volume} {7}},\ \bibinfo
  {pages} {1452} (\bibinfo {year} {2020})}\BibitemShut {NoStop}%
\bibitem [{\citenamefont {Miller}\ \emph {et~al.}(2017)\citenamefont {Miller},
  \citenamefont {Gorai}, \citenamefont {Ortiz}, \citenamefont {Goyal},
  \citenamefont {Gao}, \citenamefont {Barnett}, \citenamefont {Mason},
  \citenamefont {Snyder}, \citenamefont {Lv}, \citenamefont {Stevanovic},\ and\
  \citenamefont {Toberer}}]{miller2017capturing}%
  \BibitemOpen
  \bibfield  {author} {\bibinfo {author} {\bibfnamefont {S.~A.}\ \bibnamefont
  {Miller}}, \bibinfo {author} {\bibfnamefont {P.}~\bibnamefont {Gorai}},
  \bibinfo {author} {\bibfnamefont {B.~R.}\ \bibnamefont {Ortiz}}, \bibinfo
  {author} {\bibfnamefont {A.}~\bibnamefont {Goyal}}, \bibinfo {author}
  {\bibfnamefont {D.}~\bibnamefont {Gao}}, \bibinfo {author} {\bibfnamefont
  {S.~A.}\ \bibnamefont {Barnett}}, \bibinfo {author} {\bibfnamefont {T.~O.}\
  \bibnamefont {Mason}}, \bibinfo {author} {\bibfnamefont {G.~J.}\ \bibnamefont
  {Snyder}}, \bibinfo {author} {\bibfnamefont {Q.}~\bibnamefont {Lv}}, \bibinfo
  {author} {\bibfnamefont {V.}~\bibnamefont {Stevanovic}},\ and\ \bibinfo
  {author} {\bibfnamefont {E.~S.}\ \bibnamefont {Toberer}},\ }\bibfield
  {title} {\bibinfo {title} {Capturing anharmonicity in a lattice thermal
  conductivity model for high-throughput predictions},\ }\href@noop {}
  {\bibfield  {journal} {\bibinfo  {journal} {Chemistry of Materials}\ }\textbf
  {\bibinfo {volume} {29}},\ \bibinfo {pages} {2494} (\bibinfo {year}
  {2017})}\BibitemShut {NoStop}%
\bibitem [{\citenamefont {Toher}\ \emph {et~al.}(2014)\citenamefont {Toher},
  \citenamefont {Plata}, \citenamefont {Levy}, \citenamefont {de~Jong},
  \citenamefont {Asta}, \citenamefont {Nardelli},\ and\ \citenamefont
  {Curtarolo}}]{Toher2014PRB}%
  \BibitemOpen
  \bibfield  {author} {\bibinfo {author} {\bibfnamefont {C.}~\bibnamefont
  {Toher}}, \bibinfo {author} {\bibfnamefont {J.~J.}\ \bibnamefont {Plata}},
  \bibinfo {author} {\bibfnamefont {O.}~\bibnamefont {Levy}}, \bibinfo {author}
  {\bibfnamefont {M.}~\bibnamefont {de~Jong}}, \bibinfo {author} {\bibfnamefont
  {M.}~\bibnamefont {Asta}}, \bibinfo {author} {\bibfnamefont {M.~B.}\
  \bibnamefont {Nardelli}},\ and\ \bibinfo {author} {\bibfnamefont
  {S.}~\bibnamefont {Curtarolo}},\ }\bibfield  {title} {\bibinfo {title}
  {High-throughput computational screening of thermal conductivity, debye
  temperature, and gr\"uneisen parameter using a quasiharmonic debye model},\
  }\href {https://doi.org/10.1103/PhysRevB.90.174107} {\bibfield  {journal}
  {\bibinfo  {journal} {Physics Review B}\ }\textbf {\bibinfo {volume} {90}},\
  \bibinfo {pages} {174107} (\bibinfo {year} {2014})}\BibitemShut {NoStop}%
\bibitem [{\citenamefont {Carrete}\ \emph {et~al.}(2014)\citenamefont
  {Carrete}, \citenamefont {Li}, \citenamefont {Mingo}, \citenamefont {Wang},\
  and\ \citenamefont {Curtarolo}}]{carrete2014finding}%
  \BibitemOpen
  \bibfield  {author} {\bibinfo {author} {\bibfnamefont {J.}~\bibnamefont
  {Carrete}}, \bibinfo {author} {\bibfnamefont {W.}~\bibnamefont {Li}},
  \bibinfo {author} {\bibfnamefont {N.}~\bibnamefont {Mingo}}, \bibinfo
  {author} {\bibfnamefont {S.}~\bibnamefont {Wang}},\ and\ \bibinfo {author}
  {\bibfnamefont {S.}~\bibnamefont {Curtarolo}},\ }\bibfield  {title} {\bibinfo
  {title} {Finding unprecedentedly low-thermal-conductivity half-heusler
  semiconductors via high-throughput materials modeling},\ }\href@noop {}
  {\bibfield  {journal} {\bibinfo  {journal} {Physical Review X}\ }\textbf
  {\bibinfo {volume} {4}},\ \bibinfo {pages} {011019} (\bibinfo {year}
  {2014})}\BibitemShut {NoStop}%
\bibitem [{\citenamefont {van Roekeghem}\ \emph {et~al.}(2016)\citenamefont
  {van Roekeghem}, \citenamefont {Carrete}, \citenamefont {Oses}, \citenamefont
  {Curtarolo},\ and\ \citenamefont {Mingo}}]{van2016high}%
  \BibitemOpen
  \bibfield  {author} {\bibinfo {author} {\bibfnamefont {A.}~\bibnamefont {van
  Roekeghem}}, \bibinfo {author} {\bibfnamefont {J.}~\bibnamefont {Carrete}},
  \bibinfo {author} {\bibfnamefont {C.}~\bibnamefont {Oses}}, \bibinfo {author}
  {\bibfnamefont {S.}~\bibnamefont {Curtarolo}},\ and\ \bibinfo {author}
  {\bibfnamefont {N.}~\bibnamefont {Mingo}},\ }\bibfield  {title} {\bibinfo
  {title} {High-throughput computation of thermal conductivity of
  high-temperature solid phases: the case of oxide and fluoride perovskites},\
  }\href@noop {} {\bibfield  {journal} {\bibinfo  {journal} {Physical Review
  X}\ }\textbf {\bibinfo {volume} {6}},\ \bibinfo {pages} {041061} (\bibinfo
  {year} {2016})}\BibitemShut {NoStop}%
\bibitem [{\citenamefont {Wang}\ \emph {et~al.}(2011)\citenamefont {Wang},
  \citenamefont {Wang}, \citenamefont {Setyawan}, \citenamefont {Mingo},\ and\
  \citenamefont {Curtarolo}}]{Wang2011PRX}%
  \BibitemOpen
  \bibfield  {author} {\bibinfo {author} {\bibfnamefont {S.}~\bibnamefont
  {Wang}}, \bibinfo {author} {\bibfnamefont {Z.}~\bibnamefont {Wang}}, \bibinfo
  {author} {\bibfnamefont {W.}~\bibnamefont {Setyawan}}, \bibinfo {author}
  {\bibfnamefont {N.}~\bibnamefont {Mingo}},\ and\ \bibinfo {author}
  {\bibfnamefont {S.}~\bibnamefont {Curtarolo}},\ }\bibfield  {title} {\bibinfo
  {title} {Assessing the thermoelectric properties of sintered compounds via
  high-throughput ab-initio calculations},\ }\href
  {https://doi.org/10.1103/PhysRevX.1.021012} {\bibfield  {journal} {\bibinfo
  {journal} {Physics Review X}\ }\textbf {\bibinfo {volume} {1}},\ \bibinfo
  {pages} {021012} (\bibinfo {year} {2011})}\BibitemShut {NoStop}%
\bibitem [{\citenamefont {McGaughey}\ and\ \citenamefont
  {Kaviany}(2006)}]{mcgaughey2006phonon}%
  \BibitemOpen
  \bibfield  {author} {\bibinfo {author} {\bibfnamefont {A.~J.}\ \bibnamefont
  {McGaughey}}\ and\ \bibinfo {author} {\bibfnamefont {M.}~\bibnamefont
  {Kaviany}},\ }\bibfield  {title} {\bibinfo {title} {Phonon transport in
  molecular dynamics simulations: formulation and thermal conductivity
  prediction},\ }\href@noop {} {\bibfield  {journal} {\bibinfo  {journal}
  {Advances in Heat Transfer}\ }\textbf {\bibinfo {volume} {39}},\ \bibinfo
  {pages} {169} (\bibinfo {year} {2006})}\BibitemShut {NoStop}%
\bibitem [{\citenamefont {Schelling}\ \emph {et~al.}(2002)\citenamefont
  {Schelling}, \citenamefont {Phillpot},\ and\ \citenamefont
  {Keblinski}}]{schelling2002comparison}%
  \BibitemOpen
  \bibfield  {author} {\bibinfo {author} {\bibfnamefont {P.~K.}\ \bibnamefont
  {Schelling}}, \bibinfo {author} {\bibfnamefont {S.~R.}\ \bibnamefont
  {Phillpot}},\ and\ \bibinfo {author} {\bibfnamefont {P.}~\bibnamefont
  {Keblinski}},\ }\bibfield  {title} {\bibinfo {title} {Comparison of
  atomic-level simulation methods for computing thermal conductivity},\
  }\href@noop {} {\bibfield  {journal} {\bibinfo  {journal} {Physical Review
  B}\ }\textbf {\bibinfo {volume} {65}},\ \bibinfo {pages} {144306} (\bibinfo
  {year} {2002})}\BibitemShut {NoStop}%
\bibitem [{\citenamefont {McGaughey}\ \emph {et~al.}(2019)\citenamefont
  {McGaughey}, \citenamefont {Jain}, \citenamefont {Kim},\ and\ \citenamefont
  {Fu}}]{mcgaughey2019phonon}%
  \BibitemOpen
  \bibfield  {author} {\bibinfo {author} {\bibfnamefont {A.~J.}\ \bibnamefont
  {McGaughey}}, \bibinfo {author} {\bibfnamefont {A.}~\bibnamefont {Jain}},
  \bibinfo {author} {\bibfnamefont {H.-Y.}\ \bibnamefont {Kim}},\ and\ \bibinfo
  {author} {\bibfnamefont {B.}~\bibnamefont {Fu}},\ }\bibfield  {title}
  {\bibinfo {title} {Phonon properties and thermal conductivity from first
  principles, lattice dynamics, and the boltzmann transport equation},\
  }\href@noop {} {\bibfield  {journal} {\bibinfo  {journal} {Journal of Applied
  Physics}\ }\textbf {\bibinfo {volume} {125}},\ \bibinfo {pages} {011101}
  (\bibinfo {year} {2019})}\BibitemShut {NoStop}%
\bibitem [{\citenamefont {Gorai}\ \emph {et~al.}(2016)\citenamefont {Gorai},
  \citenamefont {Gao}, \citenamefont {Ortiz}, \citenamefont {Miller},
  \citenamefont {Barnett}, \citenamefont {Mason}, \citenamefont {Lv},
  \citenamefont {Stevanovi\'c},\ and\ \citenamefont {Toberer}}]{gorai2016}%
  \BibitemOpen
  \bibfield  {author} {\bibinfo {author} {\bibfnamefont {P.}~\bibnamefont
  {Gorai}}, \bibinfo {author} {\bibfnamefont {D.}~\bibnamefont {Gao}}, \bibinfo
  {author} {\bibfnamefont {B.}~\bibnamefont {Ortiz}}, \bibinfo {author}
  {\bibfnamefont {S.}~\bibnamefont {Miller}}, \bibinfo {author} {\bibfnamefont
  {S.}~\bibnamefont {Barnett}}, \bibinfo {author} {\bibfnamefont
  {T.}~\bibnamefont {Mason}}, \bibinfo {author} {\bibfnamefont
  {Q.}~\bibnamefont {Lv}}, \bibinfo {author} {\bibfnamefont {V.}~\bibnamefont
  {Stevanovi\'c}},\ and\ \bibinfo {author} {\bibfnamefont {E.~S.}\ \bibnamefont
  {Toberer}},\ }\bibfield  {title} {\bibinfo {title} {{TED}esign{L}ab: A
  virtual laboratory for thermoelectric material design},\ }\href@noop {}
  {\bibfield  {journal} {\bibinfo  {journal} {Computational Materials Science}\
  }\textbf {\bibinfo {volume} {112}},\ \bibinfo {pages} {368} (\bibinfo {year}
  {2016})}\BibitemShut {NoStop}%
\bibitem [{\citenamefont {Gaultois}\ \emph {et~al.}(2013)\citenamefont
  {Gaultois}, \citenamefont {Sparks}, \citenamefont {Borg}, \citenamefont
  {Seshadri}, \citenamefont {Bonificio},\ and\ \citenamefont
  {Clarke}}]{gaultois2013}%
  \BibitemOpen
  \bibfield  {author} {\bibinfo {author} {\bibfnamefont {M.~W.}\ \bibnamefont
  {Gaultois}}, \bibinfo {author} {\bibfnamefont {T.~D.}\ \bibnamefont
  {Sparks}}, \bibinfo {author} {\bibfnamefont {C.~K.~H.}\ \bibnamefont {Borg}},
  \bibinfo {author} {\bibfnamefont {R.}~\bibnamefont {Seshadri}}, \bibinfo
  {author} {\bibfnamefont {W.~D.}\ \bibnamefont {Bonificio}},\ and\ \bibinfo
  {author} {\bibfnamefont {D.~R.}\ \bibnamefont {Clarke}},\ }\bibfield  {title}
  {\bibinfo {title} {Data-driven review of thermoelectric materials:
  Performance and resource considerations},\ }\href
  {https://doi.org/10.1021/cm400893e} {\bibfield  {journal} {\bibinfo
  {journal} {Chemistry of Materials}\ }\textbf {\bibinfo {volume} {25}},\
  \bibinfo {pages} {2911} (\bibinfo {year} {2013})}\BibitemShut {NoStop}%
\bibitem [{\citenamefont {Wei}\ \emph {et~al.}(2018)\citenamefont {Wei},
  \citenamefont {Zhao}, \citenamefont {Rong},\ and\ \citenamefont
  {Bao}}]{WEI2018908}%
  \BibitemOpen
  \bibfield  {author} {\bibinfo {author} {\bibfnamefont {H.}~\bibnamefont
  {Wei}}, \bibinfo {author} {\bibfnamefont {S.}~\bibnamefont {Zhao}}, \bibinfo
  {author} {\bibfnamefont {Q.}~\bibnamefont {Rong}},\ and\ \bibinfo {author}
  {\bibfnamefont {H.}~\bibnamefont {Bao}},\ }\bibfield  {title} {\bibinfo
  {title} {Predicting the effective thermal conductivities of composite
  materials and porous media by machine learning methods},\ }\href
  {https://doi.org/https://doi.org/10.1016/j.ijheatmasstransfer.2018.08.082}
  {\bibfield  {journal} {\bibinfo  {journal} {International Journal of Heat and
  Mass Transfer}\ }\textbf {\bibinfo {volume} {127}},\ \bibinfo {pages} {908 }
  (\bibinfo {year} {2018})}\BibitemShut {NoStop}%
\bibitem [{\citenamefont {Ju}\ \emph {et~al.}(2017)\citenamefont {Ju},
  \citenamefont {Shiga}, \citenamefont {Feng}, \citenamefont {Hou},
  \citenamefont {Tsuda},\ and\ \citenamefont {Shiomi}}]{ju2017designing}%
  \BibitemOpen
  \bibfield  {author} {\bibinfo {author} {\bibfnamefont {S.}~\bibnamefont
  {Ju}}, \bibinfo {author} {\bibfnamefont {T.}~\bibnamefont {Shiga}}, \bibinfo
  {author} {\bibfnamefont {L.}~\bibnamefont {Feng}}, \bibinfo {author}
  {\bibfnamefont {Z.}~\bibnamefont {Hou}}, \bibinfo {author} {\bibfnamefont
  {K.}~\bibnamefont {Tsuda}},\ and\ \bibinfo {author} {\bibfnamefont
  {J.}~\bibnamefont {Shiomi}},\ }\bibfield  {title} {\bibinfo {title}
  {Designing nanostructures for phonon transport via bayesian optimization},\
  }\href@noop {} {\bibfield  {journal} {\bibinfo  {journal} {Physical Review
  X}\ }\textbf {\bibinfo {volume} {7}},\ \bibinfo {pages} {021024} (\bibinfo
  {year} {2017})}\BibitemShut {NoStop}%
\bibitem [{\citenamefont {Chen}\ \emph {et~al.}(2019)\citenamefont {Chen},
  \citenamefont {Tran}, \citenamefont {Batra}, \citenamefont {Kim},\ and\
  \citenamefont {Ramprasad}}]{chen2019machine}%
  \BibitemOpen
  \bibfield  {author} {\bibinfo {author} {\bibfnamefont {L.}~\bibnamefont
  {Chen}}, \bibinfo {author} {\bibfnamefont {H.}~\bibnamefont {Tran}}, \bibinfo
  {author} {\bibfnamefont {R.}~\bibnamefont {Batra}}, \bibinfo {author}
  {\bibfnamefont {C.}~\bibnamefont {Kim}},\ and\ \bibinfo {author}
  {\bibfnamefont {R.}~\bibnamefont {Ramprasad}},\ }\bibfield  {title} {\bibinfo
  {title} {Machine learning models for the lattice thermal conductivity
  prediction of inorganic materials},\ }\href@noop {} {\bibfield  {journal}
  {\bibinfo  {journal} {Computational Materials Science}\ }\textbf {\bibinfo
  {volume} {170}},\ \bibinfo {pages} {109155} (\bibinfo {year}
  {2019})}\BibitemShut {NoStop}%
\bibitem [{\citenamefont {Kautz}\ \emph {et~al.}(2019)\citenamefont {Kautz},
  \citenamefont {Hagen}, \citenamefont {Johns},\ and\ \citenamefont
  {Burkes}}]{kautz2019machine}%
  \BibitemOpen
  \bibfield  {author} {\bibinfo {author} {\bibfnamefont {E.~J.}\ \bibnamefont
  {Kautz}}, \bibinfo {author} {\bibfnamefont {A.~R.}\ \bibnamefont {Hagen}},
  \bibinfo {author} {\bibfnamefont {J.~M.}\ \bibnamefont {Johns}},\ and\
  \bibinfo {author} {\bibfnamefont {D.~E.}\ \bibnamefont {Burkes}},\ }\bibfield
   {title} {\bibinfo {title} {A machine learning approach to thermal
  conductivity modeling: A case study on irradiated uranium-molybdenum nuclear
  fuels},\ }\href@noop {} {\bibfield  {journal} {\bibinfo  {journal}
  {Computational Materials Science}\ }\textbf {\bibinfo {volume} {161}},\
  \bibinfo {pages} {107} (\bibinfo {year} {2019})}\BibitemShut {NoStop}%
\bibitem [{\citenamefont {Zhu}()}]{github_ML_kappa}%
  \BibitemOpen
  \bibfield  {author} {\bibinfo {author} {\bibfnamefont {T.}~\bibnamefont
  {Zhu}},\ }\href@noop {} {\bibinfo {title} {{Github link to data}}},\ \bibinfo
  {howpublished} {\url{https://github.com/taishanG2e/ML_kappa.git}}\BibitemShut
  {NoStop}%
\bibitem [{\citenamefont {Xie}\ and\ \citenamefont
  {Grossman}(2018)}]{xie2018crystal}%
  \BibitemOpen
  \bibfield  {author} {\bibinfo {author} {\bibfnamefont {T.}~\bibnamefont
  {Xie}}\ and\ \bibinfo {author} {\bibfnamefont {J.~C.}\ \bibnamefont
  {Grossman}},\ }\bibfield  {title} {\bibinfo {title} {Crystal graph
  convolutional neural networks for an accurate and interpretable prediction of
  material properties},\ }\href@noop {} {\bibfield  {journal} {\bibinfo
  {journal} {Physical Review Letters}\ }\textbf {\bibinfo {volume} {120}},\
  \bibinfo {pages} {145301} (\bibinfo {year} {2018})}\BibitemShut {NoStop}%
\bibitem [{\citenamefont {Rudin}(2019)}]{rudin2019stop}%
  \BibitemOpen
  \bibfield  {author} {\bibinfo {author} {\bibfnamefont {C.}~\bibnamefont
  {Rudin}},\ }\bibfield  {title} {\bibinfo {title} {Stop explaining black box
  machine learning models for high stakes decisions and use interpretable
  models instead},\ }\href@noop {} {\bibfield  {journal} {\bibinfo  {journal}
  {Nature Machine Intelligence}\ }\textbf {\bibinfo {volume} {1}},\ \bibinfo
  {pages} {206} (\bibinfo {year} {2019})}\BibitemShut {NoStop}%
\bibitem [{\citenamefont {Lee}\ \emph {et~al.}(2017)\citenamefont {Lee},
  \citenamefont {Li}, \citenamefont {Wong}, \citenamefont {Zhang},
  \citenamefont {Lai}, \citenamefont {Yu}, \citenamefont {Kong}, \citenamefont
  {Lin}, \citenamefont {Urban}, \citenamefont {Grossman} \emph
  {et~al.}}]{lee2017ultralow}%
  \BibitemOpen
  \bibfield  {author} {\bibinfo {author} {\bibfnamefont {W.}~\bibnamefont
  {Lee}}, \bibinfo {author} {\bibfnamefont {H.}~\bibnamefont {Li}}, \bibinfo
  {author} {\bibfnamefont {A.~B.}\ \bibnamefont {Wong}}, \bibinfo {author}
  {\bibfnamefont {D.}~\bibnamefont {Zhang}}, \bibinfo {author} {\bibfnamefont
  {M.}~\bibnamefont {Lai}}, \bibinfo {author} {\bibfnamefont {Y.}~\bibnamefont
  {Yu}}, \bibinfo {author} {\bibfnamefont {Q.}~\bibnamefont {Kong}}, \bibinfo
  {author} {\bibfnamefont {E.}~\bibnamefont {Lin}}, \bibinfo {author}
  {\bibfnamefont {J.~J.}\ \bibnamefont {Urban}}, \bibinfo {author}
  {\bibfnamefont {J.~C.}\ \bibnamefont {Grossman}}, \emph {et~al.},\ }\bibfield
   {title} {\bibinfo {title} {Ultralow thermal conductivity in all-inorganic
  halide perovskites},\ }\href@noop {} {\bibfield  {journal} {\bibinfo
  {journal} {Proceedings of the National Academy of Sciences}\ }\textbf
  {\bibinfo {volume} {114}},\ \bibinfo {pages} {8693} (\bibinfo {year}
  {2017})}\BibitemShut {NoStop}%
\bibitem [{\citenamefont {Kumashiro}\ \emph {et~al.}(1989)\citenamefont
  {Kumashiro}, \citenamefont {Mitsuhashi}, \citenamefont {Okaya}, \citenamefont
  {Muta}, \citenamefont {Koshiro}, \citenamefont {Takahashi},\ and\
  \citenamefont {Mirabayashi}}]{kumashiro1989thermal}%
  \BibitemOpen
  \bibfield  {author} {\bibinfo {author} {\bibfnamefont {Y.}~\bibnamefont
  {Kumashiro}}, \bibinfo {author} {\bibfnamefont {T.}~\bibnamefont
  {Mitsuhashi}}, \bibinfo {author} {\bibfnamefont {S.}~\bibnamefont {Okaya}},
  \bibinfo {author} {\bibfnamefont {F.}~\bibnamefont {Muta}}, \bibinfo {author}
  {\bibfnamefont {T.}~\bibnamefont {Koshiro}}, \bibinfo {author} {\bibfnamefont
  {Y.}~\bibnamefont {Takahashi}},\ and\ \bibinfo {author} {\bibfnamefont
  {M.}~\bibnamefont {Mirabayashi}},\ }\bibfield  {title} {\bibinfo {title}
  {Thermal conductivity of a boron phosphide single-crystal wafer up to high
  temperature},\ }\href@noop {} {\bibfield  {journal} {\bibinfo  {journal}
  {Journal of Applied Physics}\ }\textbf {\bibinfo {volume} {65}},\ \bibinfo
  {pages} {2147} (\bibinfo {year} {1989})}\BibitemShut {NoStop}%
\bibitem [{\citenamefont {Ward}\ \emph {et~al.}(2018)\citenamefont {Ward},
  \citenamefont {Dunn}, \citenamefont {Faghaninia}, \citenamefont {Zimmermann},
  \citenamefont {Bajaj}, \citenamefont {Wang}, \citenamefont {Montoya},
  \citenamefont {Chen}, \citenamefont {Bystrom}, \citenamefont {Dylla} \emph
  {et~al.}}]{ward2018matminer}%
  \BibitemOpen
  \bibfield  {author} {\bibinfo {author} {\bibfnamefont {L.}~\bibnamefont
  {Ward}}, \bibinfo {author} {\bibfnamefont {A.}~\bibnamefont {Dunn}}, \bibinfo
  {author} {\bibfnamefont {A.}~\bibnamefont {Faghaninia}}, \bibinfo {author}
  {\bibfnamefont {N.~E.}\ \bibnamefont {Zimmermann}}, \bibinfo {author}
  {\bibfnamefont {S.}~\bibnamefont {Bajaj}}, \bibinfo {author} {\bibfnamefont
  {Q.}~\bibnamefont {Wang}}, \bibinfo {author} {\bibfnamefont {J.}~\bibnamefont
  {Montoya}}, \bibinfo {author} {\bibfnamefont {J.}~\bibnamefont {Chen}},
  \bibinfo {author} {\bibfnamefont {K.}~\bibnamefont {Bystrom}}, \bibinfo
  {author} {\bibfnamefont {M.}~\bibnamefont {Dylla}}, \emph {et~al.},\
  }\bibfield  {title} {\bibinfo {title} {Matminer: An open source toolkit for
  materials data mining},\ }\href@noop {} {\bibfield  {journal} {\bibinfo
  {journal} {Computational Materials Science}\ }\textbf {\bibinfo {volume}
  {152}},\ \bibinfo {pages} {60} (\bibinfo {year} {2018})}\BibitemShut
  {NoStop}%
\bibitem [{\citenamefont {Ward}\ \emph {et~al.}(2016)\citenamefont {Ward},
  \citenamefont {Agrawal}, \citenamefont {Choudhary},\ and\ \citenamefont
  {Wolverton}}]{ward2016general}%
  \BibitemOpen
  \bibfield  {author} {\bibinfo {author} {\bibfnamefont {L.}~\bibnamefont
  {Ward}}, \bibinfo {author} {\bibfnamefont {A.}~\bibnamefont {Agrawal}},
  \bibinfo {author} {\bibfnamefont {A.}~\bibnamefont {Choudhary}},\ and\
  \bibinfo {author} {\bibfnamefont {C.}~\bibnamefont {Wolverton}},\ }\bibfield
  {title} {\bibinfo {title} {A general-purpose machine learning framework for
  predicting properties of inorganic materials},\ }\href@noop {} {\bibfield
  {journal} {\bibinfo  {journal} {npj Computational Materials}\ }\textbf
  {\bibinfo {volume} {2}},\ \bibinfo {pages} {16028} (\bibinfo {year}
  {2016})}\BibitemShut {NoStop}%
\bibitem [{\citenamefont {Togo}\ \emph {et~al.}(2015)\citenamefont {Togo},
  \citenamefont {Chaput},\ and\ \citenamefont {Tanaka}}]{phono3py}%
  \BibitemOpen
  \bibfield  {author} {\bibinfo {author} {\bibfnamefont {A.}~\bibnamefont
  {Togo}}, \bibinfo {author} {\bibfnamefont {L.}~\bibnamefont {Chaput}},\ and\
  \bibinfo {author} {\bibfnamefont {I.}~\bibnamefont {Tanaka}},\ }\bibfield
  {title} {\bibinfo {title} {Distributions of phonon lifetimes in brillouin
  zones},\ }\href {https://doi.org/10.1103/PhysRevB.91.094306} {\bibfield
  {journal} {\bibinfo  {journal} {Physics Review B}\ }\textbf {\bibinfo
  {volume} {91}},\ \bibinfo {pages} {094306} (\bibinfo {year}
  {2015})}\BibitemShut {NoStop}%
\bibitem [{\citenamefont {Lindsay}\ \emph {et~al.}(2019)\citenamefont
  {Lindsay}, \citenamefont {Katre}, \citenamefont {Cepellotti},\ and\
  \citenamefont {Mingo}}]{lindsay2019perspective}%
  \BibitemOpen
  \bibfield  {author} {\bibinfo {author} {\bibfnamefont {L.}~\bibnamefont
  {Lindsay}}, \bibinfo {author} {\bibfnamefont {A.}~\bibnamefont {Katre}},
  \bibinfo {author} {\bibfnamefont {A.}~\bibnamefont {Cepellotti}},\ and\
  \bibinfo {author} {\bibfnamefont {N.}~\bibnamefont {Mingo}},\ }\bibfield
  {title} {\bibinfo {title} {Perspective on ab initio phonon thermal
  transport},\ }\href@noop {} {\bibfield  {journal} {\bibinfo  {journal}
  {Journal of Applied Physics}\ }\textbf {\bibinfo {volume} {126}},\ \bibinfo
  {pages} {050902} (\bibinfo {year} {2019})}\BibitemShut {NoStop}%
\bibitem [{\citenamefont {Xia}\ \emph {et~al.}(2020)\citenamefont {Xia},
  \citenamefont {Pal}, \citenamefont {He}, \citenamefont
  {Ozoli{\c{n}}{\v{s}}},\ and\ \citenamefont
  {Wolverton}}]{xia2020particlelike}%
  \BibitemOpen
  \bibfield  {author} {\bibinfo {author} {\bibfnamefont {Y.}~\bibnamefont
  {Xia}}, \bibinfo {author} {\bibfnamefont {K.}~\bibnamefont {Pal}}, \bibinfo
  {author} {\bibfnamefont {J.}~\bibnamefont {He}}, \bibinfo {author}
  {\bibfnamefont {V.}~\bibnamefont {Ozoli{\c{n}}{\v{s}}}},\ and\ \bibinfo
  {author} {\bibfnamefont {C.}~\bibnamefont {Wolverton}},\ }\bibfield  {title}
  {\bibinfo {title} {Particlelike phonon propagation dominates ultralow lattice
  thermal conductivity in crystalline {Tl$_3$VSe$_4$}},\ }\href@noop {}
  {\bibfield  {journal} {\bibinfo  {journal} {Physical Review Letters}\
  }\textbf {\bibinfo {volume} {124}},\ \bibinfo {pages} {065901} (\bibinfo
  {year} {2020})}\BibitemShut {NoStop}%
\bibitem [{\citenamefont {Isaacs}\ \emph {et~al.}(2020)\citenamefont {Isaacs},
  \citenamefont {Lu},\ and\ \citenamefont {Wolverton}}]{isaacs2020inverse}%
  \BibitemOpen
  \bibfield  {author} {\bibinfo {author} {\bibfnamefont {E.~B.}\ \bibnamefont
  {Isaacs}}, \bibinfo {author} {\bibfnamefont {G.~M.}\ \bibnamefont {Lu}},\
  and\ \bibinfo {author} {\bibfnamefont {C.}~\bibnamefont {Wolverton}},\
  }\bibfield  {title} {\bibinfo {title} {Inverse design of ultralow lattice
  thermal conductivity materials via lone pair cation coordination
  environment},\ }\href@noop {} {\bibfield  {journal} {\bibinfo  {journal}
  {arXiv preprint arXiv:2004.01579}\ } (\bibinfo {year} {2020})}\BibitemShut
  {NoStop}%
\end{thebibliography}%

\end{document}